\documentclass{iau}
\usepackage{natbib}

\usepackage{graphicx}

\title[JD 11.~~Properties of IOAGs] 
{The properties of inside-out assembled galaxies at $z<0.1$  }

\author[Zewdie et al.]   
{Dejene Zewdie$^{1,2,3}$\thanks{E-mail: dejene.woldeyes@mail.udp.cl},
Mirjana Povi\'c$^{3,4}$,
Manuel Aravena$^{1}$,
Roberto J. Assef$^{1}$, and
 Asrate Gaulle$^{5}$}

\affiliation{$^1$N\'ucleo de Astronom\'ia, Universidad Diego Portales, Santiago, Chile; 
$^{2}$Department of Physics, Debre Berhan University (DBU), Debre Berhan, Ethiopia; 
$^{3}$Ethiopian Space Science and Technology Institute-AAU, Addis Ababa, Ethiopia;  
$^{4}$Instituto de Astrofisica de Andaluc\'ia (IAA-CSIC), Granada, Spain;
$^{5}$Department of Physics, Dilla University, Dilla, Ethiopia}
\pubyear{2019}
\volume{No. 356}  
\pagerange{119--126}
\jname{Nuclear Activity in Galaxies Across Cosmic Time}
\editors{M. Povi\'c, P. Marziani, J. Masegosa, H. Netzer,\\ S. H. Negu, \&
	S. B. Tessema,  eds.}
\begin{document}

\maketitle

\begin{abstract}
In this work, we study the properties of galaxies that are showing the inside-out assembly (which we call inside-out assembled galaxies; IOAGs), with the main aim to understand better their properties and morphological transformation. We analysed a sample of galaxies from the Sloan Digital Sky Survey (SDSS) Data Release 8 (DR8), with stellar masses in the range $\log M_{\star}=10.73-11.03$ $M_{\odot{}}$ at $z < 0.1$, and analyze their location in the stellar mass-SFR and the color-stellar mass diagram. We found that IOAGs  have different spectroscopic properties, most of them being classified either as AGN or composite. We found that the majority of our sources are located below the main sequence of star formation in the SFR-stellar mass diagram, and in the green valley or red sequence in the color-stellar mass diagram. We argue that IOAGs seem to correspond to the transition area where the galaxies are moving from star-forming to quiescent, and from the blue cloud to the red sequence and/or to recently quenched galaxies. 
\end{abstract}

\firstsection 
\section{Introduction}

How galaxies form and evolve through cosmic time is one of the major open questions in extragalactic astronomy and modern cosmology. In particular, we need to understand what role do galaxy mergers, AGN activity and star formation feedback play in driving the observed morphological evolution and the quenching of star formation of galaxies.

 \cite{2013Perez} showed that local galaxies with stellar masses in the range $\log$ $M=10.73 - 11.03$ $M_{\odot}$ have systematically shorter assembly times within their inner regions $(<0.5R_{50})$ when compared to that of the galaxy as a whole, contrary to lower or higher mass galaxies which show consistent assembly times at all radii. We refer to galaxies in this stellar mass range as Inside Out Assembled Galaxies, or IOAGs. 
 
In this work, we aim to characterize physical properties of IOAGs, such as morphology, star formation rates, and AGN activity, and to constrain the relation of these properties to the evolutionary state of these sources.

\section{Data and sample selection}
We select our sample of IOAGs from the Sloan Digital Sky Survey (SDSS), Data Release (DR8) MPA-JHU spectroscopic catalogue (\citealt{2011Aihara}). Following \cite{2013Perez}, we selected galaxies with the stellar mass range of $\log M_{\star}  = 10.73 - 11.03$ $ \rm{M_{\odot}}$ and redshift $z < 0.1$. We found 23816 IOAGs with spectroscopic observations from SDSS that can be classified using the NII-BPT diagram \citep{1981Baldwin, 2006Kewley}. We found that 40\% are classified as AGN (LINERs  $+$ Sy2), 40\% as composites and only 20\% as star-forming galaxies. Based on the Galaxy Zoo morphologies, 12\% of IOAGs are classified as ellipticals and 37\% as spirals. The rest (51\%) have uncertain morphologies in this catalogue.

\begin{figure*}
\begin{center}
 \includegraphics[width=15cm, height=6cm]{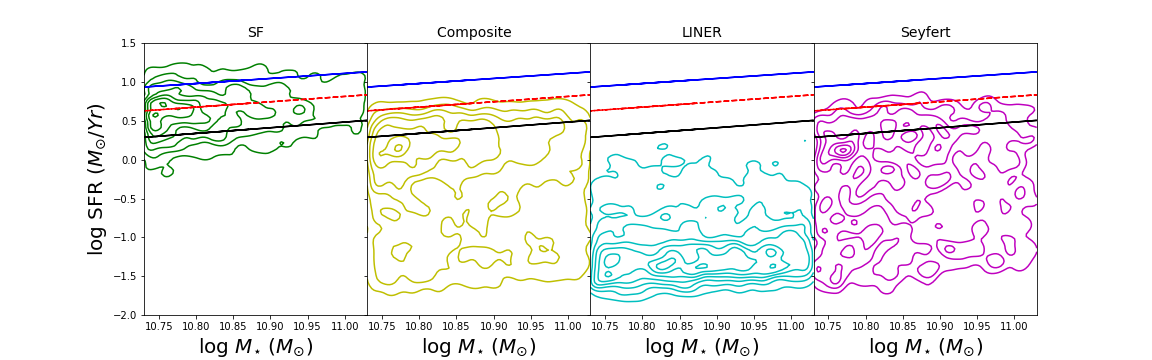}  
 \end{center}
\caption{SFR as a function of stellar mass for the IOAGs in each spectral class. A red dashed line in each panel represents the local main-sequence of star formation  from \citet{2012Whitaker}.  The blue) and black solid lines indicate the typical scatter around the MS of 0.3 dex.}
\label{fig1}
\end{figure*}

 \begin{figure*}
 \begin{center}
 \includegraphics[width=15cm, height=7cm]{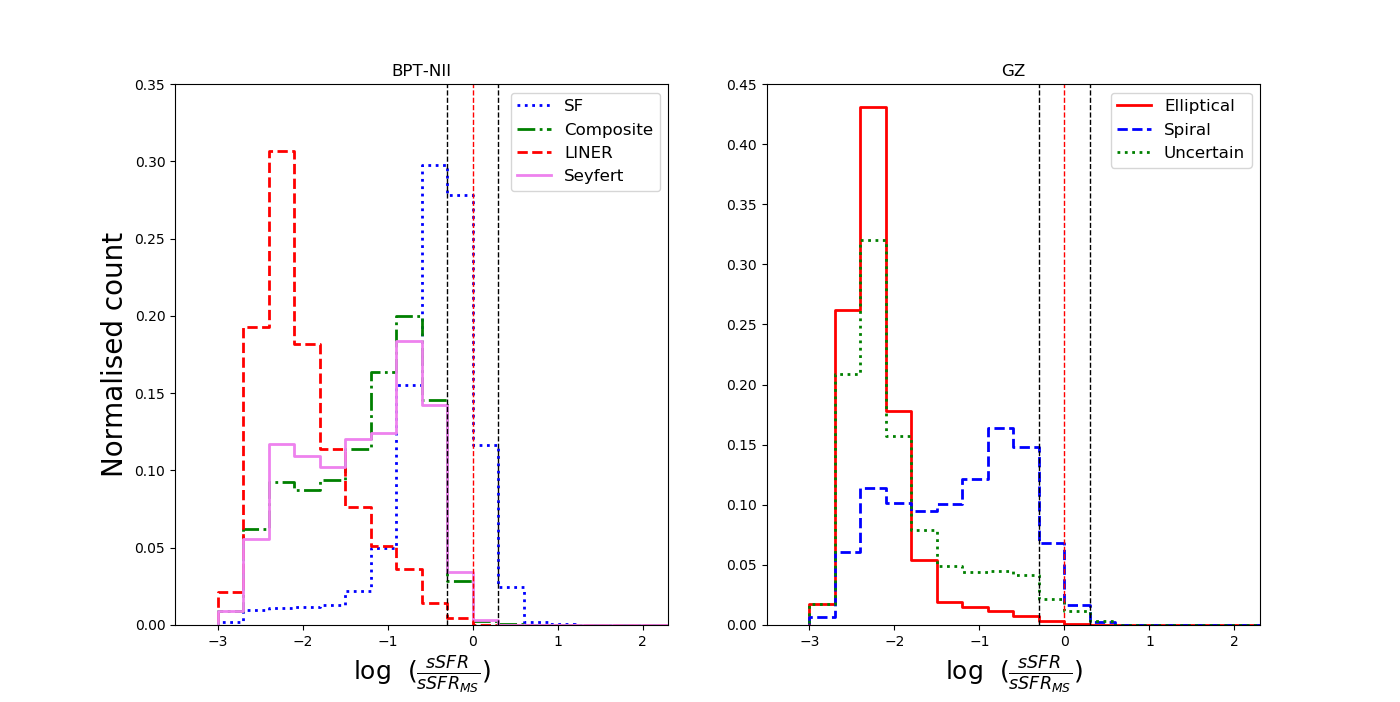} 
 \end{center}
\caption{Distribution of the sSFR, normalized to the sSFR of the MS at a given redshift and stellar mass, for the spectroscopic BPT (left)  and morphological (right) classifications. The red vertical line represents the location of the MS (at sSFR/sSFR$_{\rm{MS}}(M_*, z)=0$). The black dotted lines show the typical spread of the MS, of $\pm$0.3 dex.}
\label{fig2}
\end{figure*}

\section{Analysis and results}

Here, we have analysed the spectroscopic and morphological properties IOAGs by using the stellar masses and SFRs provided by the MPA-JHU catalogue. We found that the majority of our sources are located below the main sequence of star formation (76\%) in the SFR-stellar mass diagram. We also find the majority of IOAGs to be within the green valley (27\%) or red-sequence (49\%) in the colour-stellar mass diagram. All the details analysis and statistical results are presented in Zewdie et al. (2020, submitted).

Figure 1 shows that IOAGs spectroscopically classified as star-forming are typically consistent with the main sequence (61\%), with only 11\% (28\%) above (below) this sequence in the SFR-stellar mass diagram. Galaxies spectroscopically classified as composite, LINER or Seyfert 2 have lower SFRs, and are found to be below the main sequence and few of them are on the main sequence. With respect to their morphological classifications as ellipticals are mostly located in the red sequence, with 88\% of this population in this colour range, as expected, and 10\% in the green valley.

To visualize more clearly the number of sources of each class above on, or below the MS, we compute the specific SFR (sSFR$=$SFR/$M_*$), normalized by the sSFR of the main sequence at $z=0$ at a given stellar mass. Figure 2 shows the normalized distribution of $\delta_{MS}$ for our galaxies split by spectroscopic and morphological class. We found that the majority of the samples of IOAGs are located below the MS, some of them are on the MS and very few of the sources being in the starburst regime in both classifications.

\section{Conclusions}
We studied the morphological properties of our galaxies using the Galaxy Zoo morphological classifications, as well as their spectroscopic properties. The main results of our study are the following: 
\begin{itemize}
\item [$\star$] Most IOAGs in our spectroscopic sample may have AGN activity. 
\item [$\star$] Majority of IOAGs are located below the MS of star formation in the SFR-stellar mass diagram, and in the green valley or red sequence in the colour-stellar mass diagram.
\item [$\star$] IOAGs spectroscopically classified as SF have spiral morphologies and are in the main sequence as expected. 
\item [$\star$]Seyfert 2 and composites have spiral morphologies but quiescent SFRs, which points to the idea that the AGN could be related to their evolutionary state (Zewdie et al. 2020, submitted).
\end{itemize}

\section*{Acknowledgements}
DZ acknowledges support from the European Southern Observatory - Government of Chile Joint Committee through a grant awarded to Universidad Diego Portales. DZ and MP acknowledges financial supports from the Ethiopian Space Science and Technology Institute (ESSTI) under the Ethiopian Ministry of Innovation
and Technology (MoIT). MP acknowledges the support from the Spanish Ministry of Science,
Innovation and Universities (MICIU) through projects AYA2013-42227-P and
AYA2016-76682C3-1-P. RJA was supported by FONDECYT grant number 1191124.


\begin{thebibliography}{}
\bibitem[Baldwin et al. (1981)]{1981Baldwin}
{{Baldwin} J.~A.,  {Phillips} M.~M., and {Terlevich} R.} 1981, 
\textit{PASP}, 93, 5
\bibitem[Brinchmann et al. (2004)]{2011Aihara}
{{Brinchmann}, J., et al.,} 2004, 
\textit{MNRAS}, 351, 1151
\bibitem[Kewley et al. (2006)]{2006Kewley}
{{Kewley}, L.~J., et al.,} 2006, 
\textit{MNRAS}, 372, 961
\bibitem[P\'erez et al. (2013)]{2013Perez}
{{P{\'e}rez} E., et al.,} 2013,
\textit{ApJ}, 764, L1
\bibitem[Schawinski et al. (2014)]{2014Schawinski}
{{Schawinski} K., et al.,} 2003,
\textit{MNRAS}, 440, 889
\bibitem[Whitaker \etal\ (2012)]{2012Whitaker}
{{Whitaker} K.~E., et al.,} 2012, 
\textit{ApJ}, 754, L29






\end{thebibliography}
\end{document}